\documentclass[11pt]{article}
\usepackage{graphicx,amsfonts, amsthm, amsmath, amssymb, epic}
\theoremstyle{plain}
\newtheorem{thm}{{\bf Theorem}}[section]
\newtheorem{lemma}[thm]{{\bf Lemma}}
\newtheorem{prop}[thm]{{\bf Proposition}}
\newtheorem{cor}[thm]{{\bf Corollary}}

\newtheorem{conjecture}[thm]{{\bf Conjecture}}

\theoremstyle{remark}
\newtheorem{remark}[thm]{{\it Remark}}
\newtheorem{ex}[thm]{{\it Example}}

\def\be{\begin{eqnarray}}
\def\ee{\end{eqnarray}}
\def\ben{\begin{eqnarray*}}
\def\een{\end{eqnarray*}}
\def\ba{\begin{array}}
\def\ea{\end{array}}
\def\bp{\noindent{\it Proof. }}
\def\ep{\noindent{\hfill \fbox{}}}

\def\definition{\noindent{\bf Definition. }}
\def\pic{{\rm Pic}}

\def\noi{\noindent}
\def\nn{\nonumber}
\def\vp{\varphi}
\def\al{\alpha}

\def\chal{\alpha^{\vee}}

\def\mc{{\mathbb C}}
\def\mz{{\mathbb Z}}

\def\mr{{\mathbb R}}
\def\mpp{{\mathbb P}}

\def\nn{\nonumber}
\def\disp{\displaystyle}
\newcommand{\ol}[1]{\overline{#1}}

\newcommand{\wt}[1]{\widetilde{#1}}
\newcommand{\mapright}[1]{%
   \smash{\mathop{%
   \hbox to 1cm{\rightarrowfill}}\limits^{#1}}}
\newcommand{\mapleft}[1]{%
   \smash{\mathop{%
   \hbox to 1cm{\leftarrowfill}}\limits^{#1}}}
\newcommand{\maplleft}[2]{%
   \smash{\mathop{%
   \hbox to 1cm{\leftarrowfill}}\limits_{#1}^{#2}}}

\makeatletter       % changes [1] to 1. in bibliography
\renewcommand{\@biblabel}[1]{#1.}
\makeatother

\begin{document}

\title{Integrability of $n$-dimensional dynamical systems
of type $E_7^{(1)}$ and $E_8^{(1)}$}
\author{Tomoyuki Takenawa}
\date{}
\maketitle

\begin{center}
Faculty of Marine Technology, Tokyo University of Marine Science and 
Technology\\ 
Echujima 2-1-6, Koto-ku, Tokyo 135-8533, Japan\\
E-mail: takenawa@e.kaiyodai.ac.jp \ \ TEL \& FAX: 81 3 5245 7457
\end{center}

\begin{abstract} 

We propose an $n$-dimensional analogue of elliptic difference Painlev\'e 
equation. Some Weyl group acts on a family of rational varieties 
obtained by successive blow-ups at $m$ points in $\mpp^n(\mc)$, 
and in many cases they include the affine Weyl groups 
with symmetric Cartan matrices as subgroups. 
It is shown that the dynamical systems obtained by translations 
of these affine Weyl groups possess commuting flows and that their
degrees grow quadratically. For the $E_7^{(1)}$ and $E_8^{(1)}$
cases, existence of preserved quantities is investigated.
The elliptic difference case is also studied. 
\end{abstract} 

%\noi {\bf Mathematics Subject Classification (2000).} 14H70, (34M55, 37F, 37J)

%\noi {\bf Key words.} integrable systems, Painlev\'e equations,
%configuration space, Weyl group, algebraic degree.

\section{Introduction}

Since the introduction of the singularity confinement method by 
Grammaticos et al. \cite{grp,rgh}, the seeking of discrete integrable
systems progressed extensively. Particularly, based on the 
pioneering work by Looijenga \cite{looijenga}, Sakai classified
the relationship between discrete Painlev\'e equations and rational 
surfaces \cite{sakai}. 
On the other hand, it has become clear that
the property of singularity confinement (which is equivalent to 
the possibility of lifting maps to the sequence of 
isomorphisms of rational surfaces) 
does not guarantee integrability, and the notion of growth order of 
the algebraic degree was proposed as its reinforcement 
\cite{hv, takenawa, takenawaet}.
While the two-dimensional case has been studied in detail, there are
few studies that include the relationship between higher-dimensional integrable systems 
and higher-dimensional algebraic varieties.

\begin{figure}[ht] \label{dynkin1}
\hspace{2cm}
\begin{picture}(300,60)
\put(50,20){\line(1,0){20}}
\put(75,20){\line(1,0){10}}
\dashline{3}(85,20)(110,20)
\put(110,20){\line(1,0){10}}
\put(125,20){\line(1,0){20}}
\put(150,20){\line(1,0){20}}
\put(175,20){\line(1,0){10}}
\dashline{3}(185,20)(210,20)
\put(210,20){\line(1,0){10}}
\put(225,20){\line(1,0){20}}
\put(147.5,22.5){\line(0,1){20}}
\put(47.5,20){\circle{5}}
\put(72.5,20){\circle{5}}
\put(122.5,20){\circle{5}}
\put(147.5,20){\circle{5}}
\put(172.5,20){\circle{5}}
\put(222.5,20){\circle{5}}
\put(247.5,20){\circle{5}}
\put(147.5,45){\circle{5}}
\put(47.5,30){\makebox(0,0){}}
\put(197.5,30){\makebox(0,0){}}
\put(47.5,10){\makebox(0,0){$\al_{1}$}}
\put(72.5,10){\makebox(0,0){$\al_{2}$}}
\put(122.5,10){\makebox(0,0){$\al_{n}$}}
\put(147.5,10){\makebox(0,0){$\al_{n+1}$}}
\put(177.5,10){\makebox(0,0){$\al_{n+2}$}}
\put(227.5,10){\makebox(0,0){$\al_{m-2}$}}
\put(257.5,10){\makebox(0,0){$\al_{m-1}$}}
\put(160,45){\makebox(0,0){$\al_{0}$}}
\end{picture}
\caption[]{$W(n,m)$ Dynkin diagram}
\end{figure}
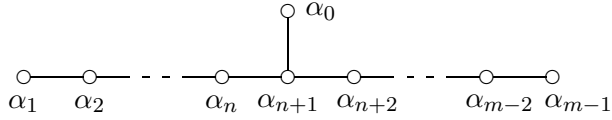

As studied by Coble \cite{coble} and Dolgachev-Ortland \cite{do},
it is known that
some Weyl groups act as pseudo-isomorphisms (isomorphisms except
sub-varieties of co-dimension 2 or higher) on
a family of rational varieties
obtained by successive blow-ups at $m$ ($m\geq n+2$) points in $\mpp^n(\mc)$.
Here, the Weyl group is given by the 
Dynkin diagram in Fig. \ref{dynkin1} and denoted by $W(n,m)$. 
The Weyl group $W(n,m)$ increases in size if $m$ increases
(Table \ref{weyltypes}).

\begin{table}[ht]\label{weyltypes}
\begin{center}

\caption{Types of Weyl groups}
\begin{tabular}{|c|c|c|c|c|c|c|c|}
\hline $n \diagdown m$ &6&7&8&9&10&11&12 \\
\hline 2 &$E_6$& $E_7$&$E_8$&$E_8^{(1)}{}^*$&
 ind.${}^*$ & ind.${}^*$ &ind.${}^*$\\
\hline 3 &$D_6$& $E_7$&$E_7^{(1)}{}^{\dag}$&ind.${}^{\dag}$&
 ind.${}^{* \dag}$ & ind.${}^{*\dag}$&ind.${}^{*\dag}$\\
\hline 4 &$A_6$& $D_7$&$E_8$&ind.${}^{\dag}$ &
 ind. ${}^{\dag}$& ind.${}^{*\dag}$ &ind.${}^{*\dag}$\\
\hline 5 &-& $A_7$&$D_8$&$E_8^{(1)}{}^{\ddag}$
& ind.${}^{\dag \ddag}$ & ind.${}^{\dag \ddag}$ &ind.${}^{*\dag \ddag}$\\ 
\hline 6 &-&  - &$A_8$&$D_9$& ind.${}^{\ddag}$ &
 ind.${}^{\dag\ddag}$ &ind.${}^{\dag \ddag}$\\ 
\hline 7 &-&  - &-&$A_9$&$D_{10}$& ind. ${}^{\ddag}$ &
 ind.${}^{\dag \ddag}$ \\ 
\hline
\end{tabular}
\end{center}
1. "ind." implies an indefinite type.\\
2. ${}^*$, ${}^{\ddag}$, and ${}^{\dag}$ imply  
that $W(E_8^{(1)})$, and
$W(E_7^{(1)})$ are included, respectively.
\end{table}

In this paper, we study the dynamical systems defined by
translations of the affine Weyl group with the symmetric Cartan matrix 
included in $W(n,m)$.
If $m\geq n+7$, $W(n,m)$ includes the affine Weyl group of type 
$E_8^{(1)}$. However, it is adequate to consider the case of $m= n+7$,
because the affine Weyl group acts trivially on
the part obtained by the blow-up at the $i$-th point for $i>n+7$.
The same argument holds true for the case of $m=n+4$ 
($n\geq 5$) or $n+5$ ($n\geq 3$). The Weyl group
$W(n,m)$ includes the affine Weyl group of type $E_8^{(1)}$
or type $E_7^{(1)}$ in each case.
In the case of $n=2, m=9$, our dynamical system coincides with the elliptic
difference Painlev\'e equation proposed by Sakai \cite{sakai},
from which discrete and continuous Painlev\'e equations are obtained by
degeneration. The case when $n=3, m=8$ was studied in \cite{takenawa3},  and 
it was revealed that this system can be reduced to the case of $n=2, m=9$.

Our dynamical systems posses commuting flows by construction,
while a general element of infinite order in the Weyl group of 
an indefinite type does not possess such flows \cite{takenawa}. 
In this paper, we show that the degrees of the systems
grow in the quadratic order, and therefore, their algebraic entropy
is zero. 
We employ the theory on the relationshpips between the algebraic degree
and the actions on cohomology groups for birational (or rational) 
dynamical systems, which has been studied by several authors for
two-dimensional autonomous cases 
(for example, Bellon \cite{bellon}, Cantat \cite{cantat}, and Diller-Favre \cite{df}).
In \cite{takenawa2}, 
the author studied two-dimensional non-autonomous
birational cases and showed that the degrees of discrete Painlev\'e equations
grow at most quadratically. 
Bedford and Kim \cite{bk} recently analyzed these relationships for
certain higher-dimensional autonomous dynamical systems 
proposed by Borkaa et al. \cite{bhm}, which occur as
a degenerated case in our systems.

Investigation of preserved quantities of our systems are divided into
(i) to solve for parameters (if it is achieved, the parameters 
are considered to be independent variables) , and (ii) to solve
for dependent variables.
Although our systems preserve some hyper-surfaces, 
it might be impossible to solve them even for their parameters.
However, it can be done if all the points of the blow-ups 
lie on an elliptic curve.
We also present a conjecture related to preserved 
quantities for dependent variables in this case.

This article is organized as follows.
In Section 2, we review the relationship between rational varieties
and groups of Cremona transformations, and we introduce $n$-dimensional 
dynamical systems associated with affine Weyl groups.
In Section 3, we calculate the algebraic degrees of the systems
and show that they grow quadratically.  
In Section 4, the existence of preserved quantities
for $E_7^{(1)}$ and $E_8^{(1)}$ cases is investigated.
In Section 5, the case where all the points of the blow-ups
lie on an elliptic curve is studied.
Section 6 is devoted to conclusions.

\section{Defining manifolds and Weyl group actions}

%\noi \ul{{\bf Variety}}

Let $m\geq n+2 $.  Let $X_{n,m}$ be the 
configuration space of ordered $m$ points in $\mpp^n$:
\ben \hspace{-1.0cm}&&
PGL(n+1) \left\backslash \left\{
\begin{pmatrix}
a_{01}&\cdots&a_{0m}\\
a_{11}&\cdots&a_{1m}\\
\vdots &\cdots&\vdots \\
a_{n1}&\cdots&a_{nm}\end{pmatrix} \left| 
\ba{l}
\mbox{the determinant of}\\
\mbox{every}\ (n+1)\times(n+1) \\
\mbox{matrix is nonzero}
\ea
\right. \right\}
\right/ (\mc^{\times})^{m} ,
\een
where $PGL(n+1)$ denotes the group of general 
projective linear transformations
of the dimension $n+1$.
$X_{n,m}$ is a quasi-projective variety of the dimension $n(m-n-2)$. 
We also consider $X_{n,m}^1\simeq X(n,m+1)$ with a natural projection 
$\pi:X_{n,m}^1\to X_{n,m}$:
\ben
\begin{pmatrix}
a_{01}&\cdots&a_{0m}&x_0\\
a_{11}&\cdots&a_{1m}&x_1\\
\vdots &\cdots&\vdots \\
a_{n1}&\cdots&a_{nm}&x_n\end{pmatrix}
&\mapsto& 
\begin{pmatrix}
a_{01}&\cdots&a_{0m}\\
a_{11}&\cdots&a_{1m}\\
\vdots &\cdots&\vdots \\
a_{n1}&\cdots&a_{nm}\end{pmatrix},
\een
where each fiber is $\mpp^n$
and $X_{n,m}$ is referred to as the parameter space.

Let $A \in X_{n,m}$, and let $X_A$
be the rational variety obtained by
successive blow-ups at the points 
$P_i=(a_{0i}:\cdots:a_{ni})$ $(i=1,2,\cdots,m)$
from $\mpp^n$. 
%Here, 
%if $P_i=P_j$ for some $i<j$, we blow-up $P_j$ after $P_i$. 
%However, this case case is not considered since it is "nodal" 
%(cf. Prop.~\ref{nodal}).
We denote the family of rational projective varieties $X_A$ ($A\in X_{n,m}$) 
as $\wt{X}_{n,m}^1$, which also has the natural fibration  
$\pi:\wt{X}_{n,m}^1 \to X_{n,m}$.

Let $E$ be the divisor class on $X_A$ of 
the total transform of a hyper-plane in $\mpp^n$, and
let $E_i$ be the exceptional divisor class generated by the blow-up 
at a point $P_i$.
The group of divisor classes of $X_A$: $\pic(X_A) \simeq 
H^1(X_A,{\cal O}^{\times})
\simeq H^2(X_A,\mz)$ (the second equivalence arises from the fact that 
$X_A$ is a rational projective variety) is described as the lattice
\be
\pic(X_A)&=&\mz E\oplus  \mz E_1 \oplus \mz E_2 \oplus \cdots \oplus 
\mz E_m.
\label{pic}
\ee
It should be noted that this cohomology group is independent of $A$, while $X_A$ is
not isomorphic to $X_{A'}$ for $A' (\neq A) \in X_{n,m}$ in general. 

Let $e \in H_2(X_A,\mz)$ be the class of a generic line in $\mpp^n$,
and let $e_i$ be the class of a generic line in the exceptional 
divisor of the blow-up at a point $P_i$.
Then, $e, e_1, e_2, \dots, e_m$ form a basis of 
$H_2(X_A,\mz)\simeq (H^2(X_A,\mz))^*$ (the equivalence is guaranteed by
the Poincar\'e duality), and the intersection numbers are given by 
$$\langle E,e\rangle =1, \langle E,e_j\rangle =0, \langle E_i, e\rangle =0, \langle E_i,e_j\rangle=-\delta_{i,j}.$$

Following Dolgachev-Ortland \cite{do}, 
we adopt the root basis $\{ \al_0,\cdots, \al_{m-1} \}
\subset H^2(X_A,\mz)$
and the co-root basis $\{ \chal_0,\cdots,\chal_{m-1} \} \subset H_2(X_A,\mz)$ as%
\ben &&\ba{ll}
\al_0= E - E_1-E_2-\cdots -E_{n+1}, & \al_i=E_i-E_{i+1} \ \ (i>0)\\
\chal_0= (n-1)e - e_1-e_2-\cdots -e_{n+1}, & \chal_i=e_i-e_{i+1} \ \ (i>0),
\ea \een
then $\langle \al_i,\chal_i\rangle=-2$ holds for any $i$, and
these root bases define the Dynkin diagram of type $T_{2,n+1,m-n-1}$
by assigning a root $\al_i$ to each vertex $\al_i$  and
connecting two distinct vertices $\al_i$ and $\al_j$ 
if $\langle \al_i,\chal_j\rangle= 1$
(in our case $\langle \al_i,\chal_j\rangle= 0 \mbox{ or }1$ for $i\neq j$) 
(Fig.~\ref{dynkin1}).

Let us define the root lattice $Q=Q(n,m) \subset H^2(X_A,\mz)$ and 
the co-root lattice $Q^{\vee}=Q^{\vee}(n,m)\subset H_2(X_A,\mz)$ as
$Q=\mz \al_0 \oplus  \mz \al_1  \oplus \cdots \oplus \mz \al_{m-1} $
and $Q^{\vee}=\mz \chal_0 \oplus  \mz \chal_1  \oplus 
\cdots \oplus \mz \chal_{m-1}$, respectively.
For each $\al_i$, the formulae
\be && \ba{rcl}
{r_{\al_i}}_*(D)&=&\disp{D+\langle D,\chal_i\rangle\al_i} \quad \mbox{for any }D\in Q \\
{r_{\al_i}}_*^{\vee}(d)&=&\disp{d+\langle \al_i,d\rangle\chal_i} \quad \mbox{for any } d 
\in Q^{\vee} \label{simref}
\ea \ee
define linear involutions (termed simple reflections) of the bi-lattice 
$(Q,Q^{\vee})$,  
and they generate the Weyl group $W$ of type $T_{2,n+1,m-n-1}$,
which we denote as $W_*(n,m)$.

%%%%%%%%%%%%%%%%%%%%%%%%%%%%%%%%%%%%%%%%%%%%%%%%%%%%%%%%%
%\noi \ul{{\bf Actions}}
 
These simple reflections correspond to certain birational transformations
on the fiber space $\pi: \wt{X}_{n,m}^1\to X_{n,m}$.
Let us define birational transformations 
$r_{i,j}$ $(1\leq i<j \leq m)$ and 
$r_{i_0,i_1,\cdots, i_n}$ $(1\leq i_0< \cdots <i_n \leq m)$ 
on the fiber space as follows:\\
$r_{i,j}$ exchanges the points $P_i$ and $P_j$
$$
r_{i,j}: 
(\ba{c|c|c|c|c|c} \cdots&{\bf a}_i& \cdots& {\bf a}_j&\cdots&{\bf x} 
\ea) \mapsto
( \ba{c|c|c|c|c|c} \cdots& {\bf a}_j& \cdots& {\bf a}_i& \cdots &{\bf x}
\ea );
$$
and $r_{i_0,i_1,\cdots, i_n}$ is the standard Cremona transformation
with respect to the points $P_{i_0}, P_{i_1}, \cdots, P_{i_n}$, 
for example, $r_{1,2,\cdots, n+1}$ is the composition of 
a projective transformation and the standard Cremona transformation
with respect to the origins $(0:\cdots:0:1:0\cdots:0)$ 
as
\ben
r_{1,2,\cdots, n+1}&:&
(A~|~{\bf x})=
(\ba{c|c|c}A_{1,\cdots, n+1}& A_{n+2,\cdots, m}& {\bf x}\ea) ~\mapsto~ 
A_{1,\cdots, n+1}^{-1} (\ba{c|c} A& {\bf x}\ea)\\
&&=:\left(\ba{c|ccc|c}
 &       &\vdots  &        &\vdots \\
I_{n+1}&\cdots & a_{ij}'& \cdots &x_i' \\
 &       &\vdots  &        &\vdots
\ea \right) ~\mapsto~ 
\left(\ba{c|ccc|c}
 &       &\vdots  &        &\vdots \\
I_{n+1}&\cdots & {a_{ij}'}^{-1}& \cdots &{x_i'}^{-1} \\
 &       &\vdots  &        &\vdots
\ea \right),
\een
where $A_{j_1,j_2,\cdots,j_k}$ denotes the $(n+1)\times k$ matrix
$({\bf a}_{j_1}~|~\cdots ~|~{\bf a}_{j_k})$, 
and $I_k$ denotes the $k \times k$ identity matrix.

\begin{remark}
The birational map 
$r_{i,j}$ cannot be defined on the fiberes of the parameters
that satisfy $P_i=P_{j}$,
in which case we blow-up at $P_j$ after at $P_i$. 
Although the points $P_i$ and $P_j$ should be 
exchanged by $r_{i,j}$, it is impossible to simultaneously preserve
the order of blow-ups.
The condition $P_i=P_j$ is equivalent to the condition 
that the divisor class  $\al_i+\cdots+\al_{j-1}=E_i-E_j 
\in H^2(X_A,\mz)$ is
effective. This situation was studied in detail by Saito and 
Umemura \cite{su}
in $n=2$ cases and its relationship with the notion of flop was revealed. 
The birational map 
$r_{1,2,\cdots,n+1}$ cannot be defined for the fibers of parameters
satisfying $\det A_{1,\cdots, n+1}=0$, which is equivalent to 
the points $P_1,P_2,\cdots,P_{n+1}$ being on the same hyper-plane
in $\mpp^n$ and the divisor class 
$\al_0=E-E_1-E_2-\cdots-E_{n+1} \in H^2(X_A,\mz)$ being effective. 
Such a hyper-surface is referred nodal one (cf. Prop.~\ref{nodal}). 
\end{remark}

\medskip

Let $w$ denote the reflection $r_{i,j}$ or $r_{i_0,i_1,\cdots, i_n}$.
The reflection $w$ acts on the parameter space $X_{n,m}$ and 
preserves the fibration 
$\pi:\wt{X}_{n,m}^1\to X_{n,m}.$ 
As previously mentioned,
$H^2(X_A,\mz)$ is independent of $A\in X_{n,m}$.
Hence, $w$ defines an action on this cohomology group.
Moreover, the induced birational map 
$w:X_A \dashrightarrow X_{w(A)}$ for generic $A\in X_{n,m}$ 
is a pseudo-isomorphism,
i.e., an isomorphism except sub-manifolds of co-dimension 2 or higher,
and the lines corresponding to the classes $e$ and $e_i$ can be chosen 
in a manner such that they do not meet the excluded part.
Since $H_2(X_A,\mz)$ is also independent of $A\in X_{n,m}$,
$w$ defines an action on this homology group and 
preserves the intersection form 
$\langle .,.\rangle : H^2(X_A,\mz)\times H_2(X_A,\mz)\to \mz$.

The birational maps $r_{i,i+1}$ and $r_{1,2,\cdots,n+1}$ correspond 
to the simple reflections ${r_{\al_i}}_*$ $(1\leq i \leq m-1)$ and 
${r_{\al_0}}_*$, respectively. Indeed, their push-forward actions on
$H^2(X_A,\mz) \to H^2(X_{w(A)},\mz)$ 
$(w=r_{i,i+1}$ or $r_{1,2,\cdots,n+1})$
and on $H_2(X_A,\mz)\to H_2(X_{w(A)},\mz)$ 
are given by the formulae:
\be && \ba{rcl}
{r_{i,i+1}}_*(D)&=&
\disp{D+\langle D,\chal_i\rangle\al_i} \\
{r_{i,i+1}}_*(d)&=&
\disp{d+\langle \al_i,d\rangle\chal_i} \\ 
{r_{1,2,\cdots,n+1}}_*(D)&=
&\disp{D+\langle D,\chal_0\rangle\al_0} \\
{r_{1,2,\cdots,n+1}}_*(d)&=&
\disp{d+\langle \al_0,d\rangle\chal_0}
\ea \label{weylact} \ee
for any $D\in H^2(X_A,\mz)$ and any $d\in H_2(X_A,\mz)$.
The formulae (\ref{weylact}) are extensions of (\ref{simref}) onto
these (co)-homology groups. 

Let $W(n,m)$ denote 
the group generated by  $r_{i,i+1}$ $(i=1,2,\cdots,m-1)$ and 
$r_{1,2,\cdots,n+1}$. It should be noted that $W(n,m)$ acts
on the $X_{n,m}$ except ``the nordal set'' 
(see Prop.~\ref{nodal}).

\begin{lemma} \label{weylid}
$W(n,m)\simeq W_*(n.m)$ holds during the correspondence
$r_{i,i+1}\simeq {r_{\al_i}}_*$ $(1\leq i \leq m-1)$ and
$r_{1,2,\cdots,n+1}\simeq {r_{\al_0}}_*$.
\end{lemma}

\bp
It is sufficient to show that for any $w_*\in W_*(n,m)$, there exists a unique
element $w\in W(n,m)$, which coincides with $w_*$ on $Q(n,m)$.
From (\ref{weylact}), 
such $w \in W(n,m)$ can be easily constructed.
Hence, we show the uniqueness of such $w$.

Let both $w,w' \in W(n,m)$ coincide with $w_* \in W_*(n,m)$ on $Q(n,m)$.
It is sufficient to prove that $w^{-1} \circ w'$ is the identity on 
$\pic(X_A)\simeq H^2(X_A,\mz)$. 

Since $w^{-1} \circ w'$ induces the identity of $Q(n,m)$ and $Q^{\vee}(n,m)$,
we can assume that $w^{-1} \circ w'$ acts on 
$H^2(X_A,\mz)$ and on $H_2(X_A,\mz)$  as
$E\mapsto E+(n+1)xK$,  $E_i \mapsto E_i+xK$, and 
$e \mapsto h+\frac{n+1}{n-1}yk$,  $e_i \mapsto e_i+yk$
for some $x,y \in \mr,\ K\in H^2(X_A,\mz)$ and $k\in H_2(X_A,\mz)$.
By $\langle \al_i,d\rangle=\langle \al_i,(w^{-1} \circ w')_*(d)\rangle$ and
$\langle D, \al_i^{\vee}\rangle=\langle (w^{-1} \circ w')_*(D),\al_i^{\vee}\rangle$
for $D=E,E_i$ and $d=e,e_i$, we have
$K=\frac{n+1}{n-1}E-E_1-E_2-\cdots-E_m$ and
$k=(n+1)e-e_1-e_2-\cdots-e_m$.
By $\langle D,d\rangle=\langle (w^{-1} \circ w')_*(D),~(w^{-1} \circ w')_*(d)\rangle$, we have
\ben
\ba{rcrcrcl}
x&+&y&+&pxy&=&0\\
x&+&\frac{n+1}{n-1}y&+&\frac{n+1}{n-1}pxy&=&0\\
(n+1)x&+&y&+&(n+1)p xy&=&0 ,
\ea
\een
where $p=\frac{(n+1)^2}{n-1}-m$;
therefore, $x=y=0$ holds. \ep \\

For a real root $\al=w_*(\al_i)$ with $w \in W(n,m)$,
we also define the reflection $r_{\al}$ as $w\circ r_{\al_i} \circ w^{-1}$.
Then, the formulae
\be && \ba{rcl}
{r_{\al}}_*(D)&=&\disp{D+\langle D,\chal\rangle\al}\\
{r_{\al}}_*(d)&=&\disp{d+\langle \al,d\rangle\chal}
\ea \label{weylact2} \ee
hold for any $D\in H^2(X_A,\mz)$ and any $d\in H_2(X_A,\mz)$.
It should be noted that both $E-E_{i_0}-E_{i_1}-\cdots- E_{i_n}$ 
and $E_i-E_j$ are real roots and their dual roots are
$(n-1)e-e_{i_0}-e_{i_1}-\cdots- e_{i_n}$ 
and $e_i-e_j$, respectively.

\medskip
We have shown the following proposition.

\begin{prop}[Coble, Dolgachev-Ortland] \label{cdo}
Let $m \geq n+2$. \\
{\rm i)} \ The actions $r_{i,i+1}$ $(1\leq i \leq m-1)$ and 
$r_{1,2,\cdots,n+1}$ 
generate the Weyl group $W(n,m)$ 
corresponding to the Dynkin diagram of 
Fig.\ref{dynkin2}.\\
{\rm ii)} \ Each element $w\in W(n,m)$ defines an action on 
$H^2(X_A,\mz)$ and $H_2(X_A,\mz)$, and 
preserves the intersection form 
$\langle .,.\rangle : H^2(X_A,\mz)\times H_2(X_A,\mz)\to \mz$.\\
{\rm iii)} \ The birational maps $r_{i,i+1}$ and $r_{1,2,\cdots,n+1}$ 
correspond 
to the simple reflections $r_{\al_i}$ $(1\leq i \leq m-1)$ and 
$r_{\al_0}$, respectively.
For a real root $\al=w(\al_i)$ and the reflection $r_{\al}=w\circ r_{\al_i} \circ w^{-1}$, 
the formulae (\ref{weylact2})
hold for any $D\in H^2(X_A,\mz)$ and any $d\in H_2(X_A,\mz)$.\\
{\rm iv)} \ Each $r_{i,j}$ or $r_{i_0,\cdots,i_n}$ is an element of $W(n,m)$.
\end{prop}
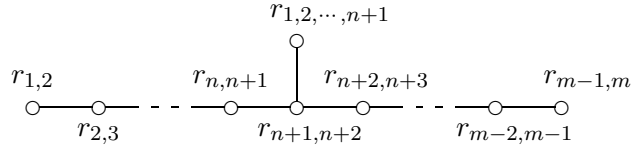
\begin{figure}[ht] \label{dynkin2}
\hspace{2cm}
\begin{picture}(300,60)
\put(50,20){\line(1,0){20}}
\put(75,20){\line(1,0){10}}
\dashline{3}(85,20)(110,20)
\put(110,20){\line(1,0){10}}
\put(125,20){\line(1,0){20}}
\put(150,20){\line(1,0){20}}
\put(175,20){\line(1,0){10}}
\dashline{3}(185,20)(210,20)
\put(210,20){\line(1,0){10}}
\put(225,20){\line(1,0){20}}
\put(147.5,22.5){\line(0,1){20}}
\put(47.5,20){\circle{5}}
\put(72.5,20){\circle{5}}
\put(122.5,20){\circle{5}}
\put(147.5,20){\circle{5}}
\put(172.5,20){\circle{5}}
\put(222.5,20){\circle{5}}
\put(247.5,20){\circle{5}}
\put(147.5,45){\circle{5}}
\put(47.5,30){\makebox(0,0){}}
\put(197.5,30){\makebox(0,0){}}
\put(47.5,30){\makebox(0,0){$r_{1,2}$}}
\put(72.5,10){\makebox(0,0){$r_{2,3}$}}
\put(122.5,30){\makebox(0,0){$r_{n,n+1}$}}
\put(152.5,10){\makebox(0,0){$r_{n+1,n+2}$}}
\put(177.5,30){\makebox(0,0){$r_{n+2,n+3}$}}
\put(230,10){\makebox(0,0){$r_{m-2,m-1}$}}
\put(257.5,30){\makebox(0,0){$r_{m-1,m}$}}
\put(160,55){\makebox(0,0){$r_{1,2,\cdots,n+1}$}}
\end{picture}
\caption[]{$W(n,m)$ Dynkin diagram}
\end{figure}

We also provide the notion of the nodal hyper-surface.  

\begin{prop}{\rm \cite{do}} \label{nodal}
The pseudo isomorphism $w:X_A \to X_{w(A)}$ 
is defined for any $w\in W(n,m)$ if and only if 
the divisor class $w_*(\al_0)$ is not effective for any $w\in W(n,m)$.
A hyper-surface in $X_A$ is referred to as nodal  
if its class is $w_*(\al_0)$ for some $w \in W(n,m)$.
\end{prop}

\bp 
Suppose $w_*(\al_0)$ is effective in $X_A$ for some $w$,
then $\al_0$ is effective in $X_{w^{-1}(A)}$ and hence  
$r_{\al_0}\circ w^{-1}:X_A \to X_{r_{\al_0}\circ w^{-1}(A)}$ is not defined.
Conversely, suppose $w_*(\al_0)$ is not effective for any $w$.
If $w:X_A \to X_{w(A)}$ is defined for some $w$, then 
$\al_0=w_*(w_*^{-1}(\al_0))$ is not effective
in $X_{w(A)}$ and hence  
$r_{\al_i}\circ w: X_A \to X_{r_{\al_i}\circ w(A)}$ is also defined.
By induction, we can define $w:X_A \to X_{w(A)}$ for any $w$. 
\ep\\

Let $N_{n,m}$ denote the set of $A \in X_{n,m}$ such that 
$X_A$ admits a nodal hyper-surface. 
From Prop.~\ref{nodal},
the Weyl group $W(n,m)$ acts on $X_{n,m}\setminus N_{n,m}$ as 
an automorphism.\\ 

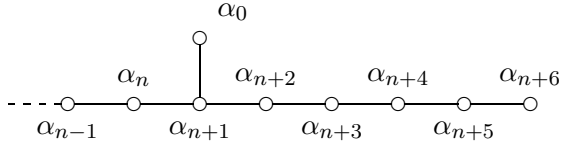
\begin{figure}[ht]
\hspace{2cm}
\begin{picture}(300,60)
\dashline{3}(25,20)(45,20)
\put(50,20){\line(1,0){20}}
\put(75,20){\line(1,0){20}}
\put(100,20){\line(1,0){20}}
\put(125,20){\line(1,0){20}}
\put(150,20){\line(1,0){20}}
\put(175,20){\line(1,0){20}}
\put(200,20){\line(1,0){20}}
\put(97.5,22.5){\line(0,1){20}}
\put(47.5,20){\circle{5}}
\put(72.5,20){\circle{5}}
\put(97.5,20){\circle{5}}
\put(122.5,20){\circle{5}}
\put(147.5,20){\circle{5}}
\put(172.5,20){\circle{5}}
\put(197.5,20){\circle{5}}
\put(222.5,20){\circle{5}}
\put(97.5,45){\circle{5}}
\put(47.5,30){\makebox(0,0){}}
\put(197.5,30){\makebox(0,0){}}
\put(47.5,10){\makebox(0,0){$\al_{n-1}$}}
\put(72.5,30){\makebox(0,0){$\al_{n}$}}
\put(97.5,10){\makebox(0,0){$\al_{n+1}$}}
\put(122.5,30){\makebox(0,0){$\al_{n+2}$}}
\put(147.5,10){\makebox(0,0){$\al_{n+3}$}}
\put(172.5,30){\makebox(0,0){$\al_{n+4}$}}
\put(197.5,10){\makebox(0,0){$\al_{n+5}$}}
\put(222.5,30){\makebox(0,0){$\al_{n+6}$}}
\put(110,55){\makebox(0,0){$\al_{0}$}}
\end{picture}
\caption[]{$E_8^{(1)}$ Dynkin diagram in $W(n,m)$}\label{e81}
\end{figure}

\begin{figure}[ht]
\hspace{2cm}
\begin{picture}(300,60)
\dashline{3}(25,20)(45,20)
\put(50,20){\line(1,0){20}}
\put(75,20){\line(1,0){20}}
\put(100,20){\line(1,0){20}}
\put(125,20){\line(1,0){20}}
\put(150,20){\line(1,0){20}}
\put(175,20){\line(1,0){20}}
\put(122.5,22.5){\line(0,1){20}}
\put(47.5,20){\circle{5}}
\put(72.5,20){\circle{5}}
\put(97.5,20){\circle{5}}
\put(122.5,20){\circle{5}}
\put(147.5,20){\circle{5}}
\put(172.5,20){\circle{5}}
\put(197.5,20){\circle{5}}
\put(122.5,45){\circle{5}}
\put(47.5,30){\makebox(0,0){}}
\put(197.5,30){\makebox(0,0){}}
\put(47.5,30){\makebox(0,0){$\al_{n-2}$}}
\put(72.5,10){\makebox(0,0){$\al_{n-1}$}}
\put(97.5,30){\makebox(0,0){$\al_{n}$}}
\put(122.5,10){\makebox(0,0){$\al_{n+1}$}}
\put(147.5,30){\makebox(0,0){$\al_{n+2}$}}
\put(172.5,10){\makebox(0,0){$\al_{n+3}$}}
\put(197.5,30){\makebox(0,0){$\al_{n+4}$}}
\put(135,55){\makebox(0,0){$\al_{0}$}}
\end{picture}
\caption[]{$E_7^{(1)}$ Dynkin diagram in $W(n,m)$}\label{e71}
\end{figure}
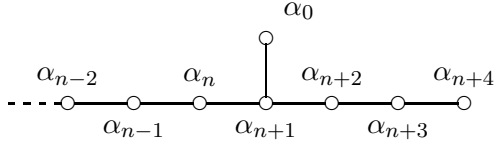

In many cases, 
the Weyl group $W(n,m)$ of an affine or indefinite type
may include affine sub-groups of the type $A^{(1)},D^{(1)},E^{(1)}$ 
(symmetric Cartan matrix type).
For example,
when $m\geq n+7$, $W(n,m)$ includes the affine Weyl group of type 
$E_8^{(1)}$, as shown in Fig.~\ref{e81}. However, it is sufficient to consider the case of $m= n+7$,
because the affine Weyl group
is generated by $r_{1,2,\cdots,n+1},$ 
$r_{n-1,n}, r_{n,n+1},\cdots, r_{n+6,n+7}$ and hence
it acts trivially on
the part obtained by the blow-up at the $i$-th point 
for $i\rangle n+7$.
When $n\geq 3, m\geq n+5$, $W(n,m)$ includes the affine Weyl group of type 
$E_7^{(1)}$ as shown in Fig.~\ref{e71}.

Next, based on Kac's book \cite{kac} (\S~6.5), 
we introduce translations on $H^2(X_A,\mz)$ and $H_2(X_A,\mz)$,
which were defined in the vector spaces $\mathfrak{h}$ and 
$\mathfrak{h}^{\vee}$ in that book, 
and we realize them as birational transformations on $\wt{X}_{n,m}^1$. 
These translations can be considered as autonomous dynamical systems on
$\wt{X}_{n,m}^1$ or non-autonomous ones on $\mpp^3$.

\noi i) 
Let $\beta_0,\beta_1,\cdots,\beta_l\in Q(n,m)$ generate
an affine Weyl group $W(R^{(1)})$ of symmetric type s.t.
$\beta_1,\cdots,\beta_l$ generate the finite Weyl group $W(R)$. 
Let $Q(R^{(1)})$ denote the root lattice generated by $\beta_i$'s 
and let $\delta$ and $\delta^{\vee}$ denote the null vector and its dual, respectively.
Here, $\langle \delta, \al_i^{\vee}\rangle=\langle \al_i,\delta^{\vee}\rangle=0$ holds.

For example, in the $E_8^{(1)}$ case, we have
$\beta_0=\al_{n+6}, \beta_i=\al_{i+n-2} \ (1\leq i \leq 7), \beta_8=\al_0$
and
\ben
\delta&=&3\beta_8+2\beta_1+4\beta_2+6\beta_3+
5\beta_4+4\beta_5+3\beta_6+2\beta_7+\beta_0\\
&=&3E-3\sum_{i=1}^{n-2}E_i-\sum_{i=n-1}^{n+7}E_i\\
\delta^{\vee}
&=&3\beta_8^{\vee}+2\beta_1^{\vee}+4\beta_2^{\vee}+6\beta_3^{\vee}+
5\beta_4^{\vee}+4\beta_5^{\vee}+3\beta_6^{\vee}
+2\beta_7^{\vee}+\beta_0^{\vee}\\
&=&3(n-1)e-3\sum_{i=1}^{n-2}e_i-\sum_{i=n-1}^{n+7}e_i.
\een

\noi ii) Let $\theta$ denote the highest root $\delta-\beta_0$
and let $\theta^{\vee}$ 
denote $\delta^{\vee}-\beta_0^{\vee}$, then 
$\theta$ is a positive root of $W(R)$ and 
thus a positive real root of $W(R^{(1)})$. 
Let $t_{\beta_0}$ denote the birational transformation 
$r_{\theta}\circ r_{\beta_0}$ on $\wt{X}_{n,m}^1$. 
From (\ref{weylact2}),
for any $D\in H^2(X_A,\mz)$ and any $d\in H_2(X_A,\mz)$,
we have 
\ben
{t_{\beta_0}}_*(D)&=&{r_{\theta}}_*(D+\langle D,\beta_0^{\vee}\rangle\beta_0)\\
&=&D-\langle D,\delta^{\vee}\rangle\al_{n+6}+\langle D,\delta^{\vee}+\beta_0^{\vee}\rangle\delta
\een
and
\ben
{t_{\beta_0}}_*(d)&=&
d-\langle \delta,d\rangle\beta_0^{\vee}+\langle \delta+\beta_0,d\rangle\delta^{\vee}.
\een

\noi iii) For any element $\beta$ of the root lattice $Q(R^{(1)})$, we
define ${t_{\beta}}_*$ as
\be
{t_{\beta}}_*(D)&=&
D-\langle D,\delta^{\vee}\rangle\beta+
\langle D, -\frac{1}{2}\langle \beta,\beta^{\vee}\rangle\delta^{\vee}+\beta^{\vee}\rangle\delta 
\label{trans2}
\ee
and
\be
{t_{\beta}}_*(d)&=&
d-\langle \delta,d\rangle\beta^{\vee}+
\langle -\frac{1}{2}\langle \beta,\beta^{\vee}\rangle\delta+\beta,d\rangle\delta^{\vee}.\label{trans3}
\ee 

\noi iv)
Additivity of ${t_{\beta}}_*$. For any 
$\al, \beta \in Q(R^{(1)})$,
\be
{t_{\al}}_* \circ {t_{\beta}}_*&=&{t_{\al+\beta}}_* \label{additive}
\ee
holds. In fact, since $\langle \al,\beta^{\vee}\rangle=\langle \beta,\al^{\vee}\rangle$ for symmetric
root systems, we have
${t_{\al}}_* \circ {t_{\beta}}_*(D)=
{t_{\al}}_*(D-\langle D,\delta^{\vee}\rangle\al+
\langle D, -\frac{1}{2}\langle \beta,\beta^{\vee}\rangle\delta^{\vee}+\beta^{\vee}\rangle\delta)
= 
D-\langle D,\delta^{\vee}\rangle(\al+\beta)+
\langle D, (-\frac{1}{2}\langle \al,\al^{\vee}\rangle-\frac{1}{2}\langle \beta,\beta^{\vee}\rangle-
\langle \beta,\al^{\vee}\rangle)\delta^{\vee}+\al^{\vee}+\beta^{\vee}\rangle\delta
={t_{\al+\beta}}*$.

\noi v) For any $w$ in $W(R^{(1)})$ and 
$\beta \in Q(R^{(1)})$,
we have 
\be
{t_{w(\beta)}}_*=w_*\circ {t_{\beta}}_*\circ w_*^{-1}. \label{trans4}
\ee
Indeed, ${w}_* \circ {t_{\beta}}_* \circ w_*^{-1}(D)=
w_*(w_*^{-1}(D)-\langle w_*^{-1}(D),\delta^{\vee}\rangle\beta
+\langle w_*^{-1}(D),\frac{1}{2}\langle \beta,\beta^{\vee}\rangle\delta^{\vee}+
\beta^{\vee}\rangle \delta)$.
Now, since $w_*(\delta)=\delta$ and $\langle .,.\rangle$ is preserved by $w$,
(\ref{trans4}) holds. 

\noi vi)  Since for any $\beta_i$  
there exists $w\in W(R^{(1)})$ such that $\beta_i=w_*(\beta_0)$,
any real root $\beta$ of $W(R^{(1)})$ can be written
as $\beta=w_*(\beta_0)$ for some $w\in W(R^{(1)})$.
Thus, for any real root $\beta$,
\ben
{t_{\beta}}_*&=&w_* \circ {t_{\beta_0}}_* \circ w_*^{-1}
\een
holds.
From (\ref{additive}), for any element 
$\beta=k_0\beta_0+k_1\beta_{n-1}+\cdots+k_l\beta_l$ 
of the root lattice $Q(R^{(1)})$, 
${t_{\beta}}_*={t_{\beta_0}}_*^{k_0}\circ {\beta_1}_*^{k_1}
\circ \cdots \circ {\beta_l}_*^{k_l}$ 
is an element of $W(R^{(1)})$. 

\noi vii)  
For $\beta_i$, 
we define the birational transformation $t_{\beta}\in W(n,m)$ 
on $\wt{X}_{n,m}^1$ by
\ben
t_{\beta_i}&=&w \circ t_{\beta_0} \circ w^{-1},
\een
where $w \in W(R^{(1)})$ such that $\beta_i=w(\beta_0)$.
%For any real root $\al=w_*(\al_{n+6})$ ($w \in W(E_8^{(1)})$) 
%we define the birational transformation $t_{\al}$ on $\wt{X}_{n,m}^1$ as
%\ben
%t_{\al}&=&w \circ t_{\al_{n+6}} \circ w^{-1},
%\een
Moreover,
for any $\beta=k_0\al_0+k_1\beta_1+\cdots+k_l\beta_l
\in Q(R^{(1)})$, 
we define the birational transformation $t_{\beta}\in W(n,m)$ 
on $\wt{X}_{n,m}^1$ by
$t_{\beta}=t_{\beta_0}^{k_0} \circ \beta_1^{k_1}\circ 
\cdots \circ \beta_l^{k_l}$. 
Now, on the level of birational transformations, 
we have $t_{\al} \circ t_{\beta}=t_{\al+\beta}$ and $t_{w_*(\al)}=w\circ
t_{\al}\circ w^{-1}$ for any $\al,\beta \in Q(R^{(1)})$ and any 
$w\in W(R^{(1)})$. These formulae are guaranteed by
the fact that the action on $H^2(X_A,\mz)$ uniquely determines
a birational transformation in $W(n,m)$ if it exists. \\

Now, we have:
\begin{prop}\label{translation}
Let $\beta \in Q(R^{(1)})$, there exists a unique birational transformation 
$t_{\beta} \in W(n,m)$ which acts on $H^2(X,\mz)$ and $H_2(X,\mz)$ as (\ref{trans2}) and 
(\ref{trans3}).
\end{prop}

\begin{ex}\label{ext}
In $E_8^{(1)}$ case, $t_{\beta_8}=t_{\al_0}$ is given by
\be && \ba{rcl}
t_{\al_0}&=&r_{1,2,\cdots,n-2,n+2,n+3,n+4}\circ 
r_{1,2,\cdots,n-2,n+5,n+6,n+7} \\
&&\circ r_{1,2,\cdots,n-2,n+2,n+3,n+4}\circ r_{1,2,\cdots,n-2,n-1,n,n+1}.
\ea \ee
In the case of $n=2, m=9$, this system coincides with the elliptic
difference Painlev\'e equation proposed by Sakai \cite{sakai}.
\end{ex}

From Prop.~\ref{translation}, the following theorem holds. 

\begin{thm}
Let $l\geq 2$ and
let $\beta_0,\beta_1,\cdots,\beta_l\in Q(n,m)$ generate
an affine Weyl group $W(R^{(1)})$ with symmetric Cartan matrix.
Let $\beta\in Q(R^{(1)})$ s.t. $\beta \not\in \mz \delta$.
There exist non-trivial flows commuting with $t_{\beta}$ on 
$\wt{X}_{n,m}^1$, i.e. \\
$\exists$ $t' \in W$ s.t.
\ben   &&(t')^k \neq (t_{\al})^l
\ (k,l \in \mz, k\neq 0) 
\mbox{ and } t_{\al} \circ t'=t'\circ t_{\al}
\een
\end{thm}

%%%%%%%%%%%%%%%%%%%%%%%%%%%%%%%%%%%%%

\section{Algebraic degrees of dynamical systems}

Let $X,Y,Z$ be varieties obtained by blow-ups from $\mpp^n$,
and let $\vp:X \dashrightarrow Y$ be a dominant rational map,
i.e., the closure of $\vp(X)$ coincides with $Y$.
Let $\vp^*$ denotes the pull-buck of $\vp$: $\vp^*:\pic(Y)\to \pic(X)$.
We define the algebraic degree of $\vp$: $\deg{\vp}$ as the degree of
the deduced rational map $\vp:\mpp^n \dashrightarrow \mpp^n$.

\begin{prop}
The degree of $\vp$ coincides with the coefficient of $E$ of $\vp^*(E)$,
where $E$ denotes the class of a hyper-plane in $\mpp^n$.
\end{prop}

\bp
Write $\vp:\mpp^n \dashrightarrow \mpp^n$ as 
$(f_0({\bf x}):f_1({\bf x}):\cdots:f_n({\bf x}))$
by polynomials $f_i$, where $f_i$'s are simplified if possible. 
The class $E \in \pic(Y)$  corresponds to a hyper-plane in $Y$:
$a_0 y_0' +a_1 y_1'+\cdots a_n y_n'=0$ and the class $\vp^*(E)$ corresponds
to the hyper-surface in $X$: 
$a_0f_0({\bf x})+a_1f_1({\bf x})+\cdots + a_nf_n({\bf x})$.
Hence the coefficient of $E$ of $\vp^*(E)$ is the degree of
the polynomials $f_i$.
\ep\\

\begin{remark}
Let $X_i$ $(i \in \mz)$ be varieties obtained by blow-ups from $\mpp^n$, 
and let $\vp_i:X_{i} \dashrightarrow X_{i+1}$ be a dominant rational map.
\be 
&& \liminf_{k\to \infty} \frac{1}{k} 
\log \deg (\vp_{k-1}\circ \vp_{k-2} \circ \cdot \circ \vp_0)
\ee
is referred to as the algebraic entropy of the dynamical systems $\{\vp_i\}$.
This notion is closely related to other entropies.
Indeed, by Gromov-Yomdin's theorem \cite{gromov,yomdin}, 
for a surjective morphism $f$ from a K\"ahler complex manifold $X$ to itself,
the topological entropy of $f$ equals $\log \lambda(f)$, where
$\lambda(f)$ is the spectral radius of 
$f^*:H^{\bullet}(X,\mr)\to H^{\bullet}(X,\mr)$. On the other hand,
its algebraic degree is given (or defined) by the action on 
a linear subspace $H^{2}(X,\mr)$.\\
\end{remark}

\definition
For a morphism $\psi:X\to Y$, let $\psi_c^*: {\cal P}(Y) \to {\cal P}(X)$ 
denote the set-correspondence
defined by the pull-buck, i.e. for $V \subset Y$, $\psi_c^*(V)$ is defined by
$$\psi_c^*(V)=\{x \in X~;~ \psi(x)\in V \}.$$
Let $\vp: X \dashrightarrow Y$ be a dominant rational map.  
Let $X'$ be the varieties obtained by successive blow-ups
$\pi: X' \to X$, 
which eliminates the indeterminacy of $\vp$, and 
let $\vp':X'\to Y$  be the map lifted from $\vp$ (Fig.~\ref{lift}).  
The set correspondence $\vp_c{}_*: {\cal P}(X) \to {\cal P}(Y)$ is
defined by
$$\vp_c{}_*:= \vp' \circ \pi_c^*$$
($=(\vp^{-1})_c^* $ if $\vp$ is birational)
and the  set correspondence $\vp_c^*: {\cal P}(Y) \to {\cal P}(X)$ is
defined by
$$\vp^*:=\pi \circ \vp'{}_c^* ~.$$

\begin{figure}[ht] 
\hspace{2cm}
\begin{picture}(250,60)
\put(120,55){\mbox{$X'$}}
\put(120,20){\mbox{$X$}}
\put(165,20){\mbox{$Y$}}
\put(140,40){\mbox{$\searrow$}}
\put(148,48){\mbox{$\vp'$}}
\put(122.5,37.5){\mbox{$\downarrow$}}
\put(112.5,35){\mbox{$\pi$}}
\put(135,20){\mbox{$\dashrightarrow$}}
\put(140,10){\mbox{$\vp$}}
\end{picture}
\caption[]{Lift of $\vp$}\label{lift}
\end{figure}

\definition 
We define the indeterminate component set ${\cal I}(\vp)$ and 
the critical value component
set ${\cal C}(\vp)$ for a dominant rational map $\vp:X \dashrightarrow Y$ as
\ben && \ba{rcll}
{\cal I}(\vp)&:=&\{L \in X;& \mbox{$\exists S \subset Y$: 
irreducible hyper-surface s.t.
$L$ is an irreducible}\\
&&& \mbox{component of $\vp_c^*(S)$ and $\dim L < n-1$} \} \\
{\cal C}(\vp)&:=&\{L \in Y;& \mbox{$\exists S \subset X$: 
irreducible hyper-surface s.t.
$L$ is an irreducible}\\
&&& \mbox{component of $\vp_c{}_*(S)$ and $\dim L <n-1$} \}. 
\ea \een

\begin{prop} \label{pr2} Let $f:X\dashrightarrow Y$ and 
$g:Y \dashrightarrow Z$ be dominant rational maps.
Then 
$$ (g\circ f)^* (D) = f^*\circ g^* (D)$$   holds if
and only if
${\cal C}(f)\cap {\cal I}(g)=\emptyset$.
\end{prop}

\bp
Without loss of generality, we can assume that $D$
is an irreducible hyper-surface. 
There exist irreducible hyper-surfaces $D_1,D_2,\cdots,D_m \subset Y$, 
subvarieties $L_1,L_2,\cdots, L_l \in {\cal I}(g)$,
and positive integers $r_1,r_2,\cdots,r_m$
such that 
$g_c^*(D)=D_1 + D_2+\cdots + D_m + L_1+ \cdots
+L_l$ (``$+$" implies set sum) and
$g^*(D)= r_1D_1 + r_2D_2 +\cdots + r_mD_m$. 
Similarly, for each $D_i$, there exist
irreducible hyper-surfaces
$D_{i,1},D_{i,2},\cdots,D_{i,m_i} \subset X $,
subvarieties $L_{i,1},L_{i,2},\cdots, L_{i,l_i} \in {\cal I}(f)$,
and positive integers $r_{i,1},r_{i,2},\cdots,r_{i,m_i}$
such that 
$f_c^*(D_i)=D_{i,1} + \cdots +D_{i,m_i} + L_{i,1}+ 
\cdots + L_{i,l_i}$ and 
$f^*(D_i)=r_{i,1}D_{i,1} + \cdots + r_{i,m_i}D_{i,m_i}$. Hence, we have
$$f^*\circ g^* (D)=\sum_{i=1}^m \sum_{j=1}^{m_i} r_{i,j}D_{i,j}.$$
Now, it can be easily observed that 
$(g\circ f)^* (D)-f^*\circ g^* (D) \geq 0$ holds, and the equality holds
iff $L_i \in {\cal C}(f)$ for some $i$.
\ep \\

The following proposition immediately follows from Prop.~\ref{pr2}.
\begin{prop}
Let $X_i$ $(i \in \mz)$ be varieties obtained by blow-ups from $\mpp^n$, and let $\vp_i:X_{i} \dashrightarrow X_{i+1}$ be a dominant rational map.
Then, 
\be
(\vp_{i+j} \circ \cdots \circ \vp_i)^*=
\vp_i^* \circ \cdots \circ \vp_{i+j}^* \label{as}
\ee
for any $i,j$
if and only if  
${\cal C}(\vp_{i+j-1} \circ \cdots \circ \vp_{i})\cap {\cal I}(\vp_{i+j})$ 
holds for any $i,j$.
\end{prop}
A sequence of dominant rational maps satisfying condition 
(\ref{as}) for any $i,j
\geq 0$ is referred to as ``analytically stable" in \cite{sibony} 
(cf. \cite{df})
\footnote{In \cite{bk},  Bedford and Kim 
referred to this notion ``$(1,1)$-regularity". Here, the term ``$(1,1)$" 
arises from the fact that 
$\disp{c_1(\pic(X))= c_1(H^1(X,{\cal O}^{\times})) 
\simeq H^{1,1}(X,\mz):=H^{1,1}(X)\cap H^2(X,\mz)}$, where $c_1$ is 
the 1st Chern map.}

\begin{cor}
A sequence of pseudo-isomorphisms is analytically stable. 
\end{cor}

\begin{ex}
Set $n=4$ in example~\ref{ext}. Further, an element of $\pic(X_A)$ is 
written as
$bE+b_1E_1+b_2E_2+\cdots+b_{11}E_{11}$.
For the translation $t_{\al_0}$ in example~\ref{ext}, the induced map
$t_{\al_0}^*: \pic(X_{t_{\al_0}(A)}) \to \pic(X_{A})$ is described as a
linear transformation for the vector $[b,b_1,b_2,\cdots,b_{11}]^t$ as
{\scriptsize
\ben 
t_{\al_0}^*
&=&\left[ \ba{rrrrrrrrrrrr}
28& 9& 9& 1& 1& 1& 4& 4& 4& 4& 4& 4\\
-27& -8& -9& -1& -1& -1& -4& -4& -4&-4& -4& -4\\
-27& -9& -8& -1& -1& -1& -4& -4& -4& -4& -4& -4\\
-15& -5& -5& 0& -1& -1& -2& -2& -2& -2& -2& -2\\
-15& -5& -5& -1& 0& -1& -2& -2& -2& -2& -2& -2\\
-15& -5& -5& -1& -1& 0& -2& -2& -2& -2& -2& -2\\
-6& -2& -2& 0& 0& 0& 0& -1& -1& -1& -1& -1\\
-6& -2& -2& 0& 0& 0& -1& 0& -1& -1& -1& -1\\
-6& -2& -2& 0& 0& 0& -1& -1& 0& -1& -1& -1\\
-6& -2& -2& 0& 0& 0& -1& -1& -1& 0& -1& -1\\
-6& -2& -2& 0& 0& 0& -1& -1& -1& -1& 0& -1\\
-6& -2& -2& 0& 0& 0& -1& -1& -1& -1& -1& 0
\ea \right] .
\een }
Since the cells of the Jordan normal form are
$$1,~1,~1,~1,~1,~1,~1,~1,~1,~\left[ \ba{ccc}1&1&0\\0&1&1\\0&0&1\ea \right],$$
the elements of $(t_{\al_0}^l)^*$ grow quadratically with respect to $l$.
\end{ex}

The following theorem generalizes this example.

\begin{thm}
Let $\beta_0,\beta_1,\cdots,\beta_l\in Q(n,m)$ generate
an affine Weyl group $W(R^{(1)})$ of the symmetric type. 
Let $\beta \in Q(R^{(1)})$ be not in $\mz \delta$,
and let $t_{\beta}$ be the translation defined as in the previous section.
There exists $\lim_{l \to \infty} (\deg t_{\beta}^l)/l^2 > 0$,
and therefore, the algebraic entropy of $t_{\beta}$: 
$\lim_{l\to \infty}\frac{1}{l}\log \deg t_{\beta}^l$ is $0$.
\end{thm}

\begin{remark}
From (\ref{additive}), if $\beta\in \mz \delta$, then $t_{\beta}$ is the identity.\\
\end{remark}

\bp
From Prop.~\ref{translation}, we have  
\ben(t_{\beta}^l)^*(E)&=&t_{l\beta}^*(E)
={t_{-l\beta}}_*(E)\\
&=&
E+l\langle E,\delta^{\vee}\rangle\beta-
\langle E, \frac{1}{2}\langle \beta,\beta^{\vee}\rangle l^2\delta^{\vee}+l\beta^{\vee}\rangle\delta\\
&=&
E+(n-1)l\beta-
 \left(\frac{1}{2}\langle \beta,\beta^{\vee}\rangle(n-1)l^2+ \langle E,\beta^{\vee}\rangle l\right)\delta;
\een
therefore, the coefficient of $E$ is 
in the order $l^2$ except the case where $\langle \beta,\beta^{\vee}\rangle=0 
\Longleftrightarrow \beta\in \mz \delta$.
\ep\\

%%%%%%%%%%%%%%%%%%%%%%%%%%%%%%%%%%%%%%%%%

\section{Preserved divisor classes}

In this section, for the $E_7^{(1)}$ and
$E_8^{(1)}$ cases, we investigate the existence of quantities 
 on the fibers $X_A$ 
preserved by the translation $t_{\beta}:X_A\to X_{t_{\beta}(A)}$.

In order to find preserved quantities, we have to chose coordinates
of $X_{n,m}^1$ so that the preserved hyper-surfaces are fixed by
the time evolution. It should be noted that the class of the proper transform  of a hyper-surface in $\mpp^3$ is
$k_0 E- \sum_{i=1}^m k_iE_i \in \pic(X_A)$, where 
$k_0\geq 1$ is the degree of the surface and $k_i\geq 0$ ($i=1,2,\cdots,m$)
is its multiplicity at the point $P_i$.

Let $m \geq n+7$.
We set $\beta_0,\beta_1,\cdots,\beta_l\in Q(n,m)$ in a manner such that
they generate the affine Weyl group $W(E_8^{(1)})$, as described in Section~2.
The result for the $E_7^{(1)}$ case is presented 
at the end of this section. 

we show the following theorem.

\begin{thm}\label{prediv}
Let $\beta \in Q(E_8^{(1)})$ be not in $\mz \delta$.
For a generic parameter $A\in X(n,m)$, $\delta$ is 
(a unique effective class (up to a constant multiple)) preserved by
the dynamical system $t_{\beta}$.
\end{thm} 

\begin{remark}
Unfortunately, uniqueness is still a conjecture.
\end{remark}

The parameters $A\in X(n,m)$ can be normalized as
\be 
\left( 
\ba{ccc|cccc|ccc} 
&&&b_{11}&b_{12}&\cdots&b_{19} &c_{0,n+8}&\cdots&c_{0,m}\\
&I_{n-2}&&\vdots&\vdots&\vdots&\vdots&&&\\
&&&b_{n-2,1}&b_{n-2,2}&\cdots&b_{n-2,9}&\vdots&\vdots&\vdots\\ \hline
&&&1&1&\cdots&1&\vdots&\vdots&\vdots\\
&0&&\wp(u_1)&\wp(u_2)&\cdots&\wp(u_9) &&&\\
&&&\wp'(u_1)&\wp'(u_2)&\cdots&\wp'(u_9)&c_{n,n+8}&\cdots&c_{n,m}
\ea
\right),
\ee
where $(b_{ij})_{1\leq i \leq n-2,\ 1\leq j \leq 4}$ can be fixed as
\ben
(b_{ij})_{1\leq i \leq n-2,\ 1\leq j \leq 4}&=&
\begin{pmatrix}
0&0&0&1\\
\vdots&\vdots&\vdots&\vdots\\
0&0&0&1\\
\end{pmatrix}
\een
and $\wp(u)$ is the Wierstrass $\wp$ function of periods $(1,\tau)$.
The linear system of $\delta$ is then given by the hyper-surface
$$x_{n-1}x_{n+1}^2=x_n^3-g_2x_{n-1}^2x_n-g_3x_{n-1}^3,$$
where $g_2$ and $g_3$ are constants determined by the periods.

The action of $W(E_8^{(1)})$ on $(x_{n-2}:x_{n-1}:x_n)\in \mpp^3$
is the same as that with Sakai's elliptic difference Painlev\'e equation
\cite{sakai}; hence, we arrived at the following corollary.

\begin{cor}
The period $\tau$ is preserved by $W(E_8^{(1)})$, and the
actions on parameters $u_i$ are linear transformations, which are described as follows:\\
{\rm i)} $r_{\beta_8}$ acts on $u_i$ and on $u\ (\neq u_i)$ as\\
$u_i\mapsto u_i-\frac{2}{3}(u_1+u_2+u_3)$ $(1\leq i\leq 3)$, \
$u_i\mapsto u_i+\frac{1}{3}(u_1+u_2+u_3)$ $(4\leq i\leq 9)$,\
$u\ (\neq u_i) \mapsto u+\frac{1}{3}(u_1+u_2+u_3)$;\\
{\rm ii)}  $r_{\beta_j}$ $(1\leq j\leq 7)$ acts as\\
$u_i\mapsto u_i $ $(i\neq j,j+1)$, \
$u_j\mapsto u_{j+1}$ , \
$u_{j+1}\mapsto u_j $ \
$u\ (\neq u_i) \mapsto u$;\\
{\rm iii)} $r_{\beta_0}$  acts as\\
$u_i\mapsto u_i $ $(i\neq 8,9)$, \
$u_8\mapsto u_9$ , \
$u_9\mapsto u_8$,\
$u(\neq u_i) \mapsto u$.\\
\end{cor}

\noi {\it Proof of the uniqueness of Th.~\ref{prediv}}

Fix $\beta \in Q(E_8^{(1)})$
and suppose $D\in \pic(X_A)$ is preserved by $t_\beta$.
Since the coefficients of $\beta$ and $\delta$ in formula 
(\ref{trans2}) should be zero,
$D$ is in the sub-lattice spanned by
$\al_i (1\leq i \leq n-3, n-1\leq i \leq n+6)$, $\delta$,
and the canonical divisor class $K_{X_A}=-(n+1)E+(n-1)\sum_{i=1}^m E_i$.

The proof is based on the following elementary conjecture. 
The author can prove only the case of $n=2,3$ in an elementary way.

\begin{conjecture}\label{lindim1}
Let the points $P_i$ ($1\leq i \leq m$) be in a generic position.
The dimension of the linear system $|k_0E-\sum_{i=1}^m k_iE_i|$ is given by
(negative dimension implies the empty set in this case):
\be &&{}_{n+1}H_{k_0}-1+\sum_{s=1}^m 
(-1)^s \sum_{1\leq i_1 <i_2<\cdots <i_s \leq m} 
\max\{ {}_{n+1}H_{k_{i_1}+k_{i_2}+\cdots+k_{i_s}-(s-1)k_0-s},\ 0\},
\ee
where $\disp {}_uH_v= \begin{pmatrix}u+v-1\\v 
\end{pmatrix}$ denotes the repeated combination.
\end{conjecture}

\begin{lemma}\label{lindim2}
Let $k_i>k_j\geq 0$.
For divisors
$$D=k_0E-\sum_{i=1}^m k_iE_i$$ and 
$$D'=k_0E-\left(\sum_{i=1}^m k_iE_i\right)-E_i+E_j,$$
the inequality $\dim(|D|)\leq \dim (|D'|)$ holds, where the equality
holds if and only if $k_i=k_j+1$.
\end{lemma}

\bp Without loss of generality, we can assume $i=1, j=2$. 
We define function $f_d(k_1,k_2,x,y)$ ($d=1,2,3,\cdots$ and $x,y\in \mr$)  
recursively as  
$$f_1(k_1,k_2,x,y)=
-\left(\begin{pmatrix}n+k_1+x+y-1\\n\end{pmatrix}\vee 0\right)
-\left(\begin{pmatrix}n+k_2+x-y-1\\n\end{pmatrix}\vee 0\right),$$
$$f_d(k_1,k_2,x,y)=f_{d-1}(k_1,k_2,x,y)-f_{d-1}(k_1,k_2,x+k_{d+1},y).$$
It is sufficient to show that $f_d(k_1,k_2,x,y)$ decreases
with respect to $y$. By induction for $d$, it is easy to show that 
$\disp \frac{\partial^{s+t} f_d}{\partial x^s \partial y^t}\leq 0$
for $s = 0,1,2,\cdots $ and $t=0,1$.
\ep\\

From Conj.~\ref{lindim1}, $\dim(|z\delta|)=0$ and 
 $\dim|(-zK_{\wt{X}_A}|)<0$  hold for $z>0 \in \mr$,
and thus, by Lemma~\ref{lindim2}$, \delta$ (up to a constant multiple) 
is the unique effective class preserved by $t_{\beta}$. \\
\ep \ (Uniquness of the Th.~\ref{prediv}) \\

\noi {\it Results for the $E_7^{(1)}$ case} \\

Let $m\geq n+5$.
We set the root basis $\beta_i$ as
$\beta_7=\al_0$, $\beta_i=\al_{n-3+i}$ $(1\leq i \leq 6)$,
and $\beta_0=\al_{n+4}$. Further, we also set
the null root $\delta$ as
$\delta=2\beta_7+\beta_1+2\beta_2+3\beta_3+4\beta_4+3\beta_5+2\beta_6+\beta_0$.
The translation $t_{\beta}$ is defined by (\ref{trans2}). 
For generic parameters,  $\delta$ is
preserved by $t_{\beta}$ $(\beta \not\in \mz\delta)$.

The parameters $A\in X(n,m)$ can be normalized as
\be 
\left( 
\ba{ccc|cccc|ccc} 
&&&b_{11}&b_{12}&\cdots&b_{18} &c_{0,n+6}&\cdots&c_{0,m}\\
&I_{n-2}&&\vdots&\vdots&\vdots&\vdots&&&\\
&&&b_{n-2,1}&b_{n-2,2}&\cdots&b_{n-2,8}&\vdots&\vdots&\vdots\\ \hline
&&&1&1&\cdots&1&\vdots&\vdots&\vdots\\
&0&&\wp(u_1)&\wp(u_2)&\cdots&\wp(u_8) &&&\\
&&&\wp'(u_1)&\wp'(u_2)&\cdots&\wp'(u_8)&&&\\
&&&\wp(u_1)^2&\wp(u_2)^2&\cdots&\wp(u_8)^2&c_{n,n+6}&\cdots&c_{n,m}\\
\ea
\right),
\ee
where $(b_{ij})_{1\leq i \leq n-3,\ 1\leq j \leq 4}$ can be fixed as
\ben
(b_{ij})_{1\leq i \leq n-3,\ 1\leq j \leq 5}&=&
\begin{pmatrix}
0&0&0&0&1\\
\vdots&\vdots&\vdots&\vdots&\vdots\\
0&0&0&0&1\\
\end{pmatrix}.
\een
The corresponding divisor of $\delta$ is given by the pencil of the hyper-surfaces
$$
C_1(x_{n-1}^2-x_{n-2}x_n+g_2x_{n-2}x_{n-3}+g_3x_{n-3}^2)+
C_2(x_{n-2}^2-x_{n-3}x_n)=0,$$
where $C_1$ and $C_2$ are constants s.t. $(C_1,C_2) \in \mpp^1$.

The action of $W(E_7^{(1)})$ on $(x_{n-3}:x_{n-2}:x_{n-1}:x_n)\in \mpp^3$ 
is the same with that on $X(3,8+1)$, which was studied in \cite{takenawa3}. 
From the result of \cite{takenawa3}, the action of $W(E_7^{(1)})$ on
$\mpp^n$
preserves each hyper-surface, which is independent of the parameters $A$, and hence it also preserves $\tau$ and $(C_1:C_2)$.
Moreover, the action can be reduced to 
the action on the lower dimensional rational variety  
$(\mpp^1\times \mpp^1)\times \mc^{n-3}$ 
through Segr\'e embedding
$\mpp^1\times \mpp^1 \to \mpp^3: (x_0:x_1, y_0:y_1)\mapsto 
(x_0y_0:x_0y_1:x_1y_0:x_1y_1)$.

\section{Elliptic difference case}

Since the time evolution of the parameters $A$ may not
be solved in general as in the previous section, the systems should not be considered as 
$n$-dimensional but $n+mn-((n+1)^2-1)$
$=n+ (\mbox{the freedom of $m$ points}) - (\mbox{the freedom of } PGL(n))$
-dimensional. On the other hand, Kajiwara {\it et al.} \cite{kmnoy}
has proposed a birational representation of the Weyl group $W(n,m)$,
in which all the points of blow-ups lie on a certain elliptic curve in $\mpp^n$,
and the actions on the parameters can be written as linear transformations
on a torus \footnote{This calculation was carried out in a rather heuristic manner. We can recover it in an algebro geometric way for the elliptic curves of 
degree $n+1$ \cite{et}.}.  
In this section, we show that the time evolution
of the parameters can be solved in the case where the points of 
the blow ups are 
parameterized  as in Kajiwara {\it et al.}. Further, we represent
the dynamical systems explicitly.

Suppose that the points $P_i$ $(i=1,2,\cdots,m)$ are on the curve
\be 
\left\{P(z)= \left(\frac{[\lambda+z_1-z]}{[z_1-z]}:\cdots:  
\frac{[\lambda+z_{n+1}-z]}{[z_{n+1}-z]} \right);~z \in \mc \right\} ,
\label{curve}
\ee
where $[z]=z,~ \sin z$ or $\theta_{11}(z)$: a theta function whose
zero points are $\mz+ \mz \tau$ with the order 1, \ 
$\lambda=z_0-z_1-z_2-\cdots-z_{n+1} \in \mc$, 
and the point $P_i$ corresponds to $z=z_i\in \mc$.
The following proposition is due to \cite{kmnoy}.

\begin{prop}[Kajiwara {\it et al.}] \label{actr}
The birational maps $r_{i,i+1}$  
act on the parameter space $(z_1, \cdots, z_m, \lambda)$ as
\be
r_{i,i+1} &:&  (z_1,\cdots,z_{i},z_{i+1},\cdots,z_{n+1};\lambda)\mapsto 
(z_1,\cdots,z_{i+1},z_i,\cdots,z_{n+1};\ol{\lambda}),  \label{r12}
\ee
where
\ben
\ol{\lambda}&=& \left\{ \ba{ll}
\lambda & ( i\neq n+1)\\
\lambda+z_{n+1}-z_{n+2} & (i=n+1)
\ea \right. ,
\een
and they act on $\mpp^n$ as
\be
r_{i,i+1}: {\bf x} &\mapsto& \ol{{\bf x}},
\ee
where 
\ben
\ol{{\bf x}}
=(x_0:\cdots:x_{i}:x_{i-1}:\cdots:x_{n}) && (\mbox{{\rm if}}\quad 
 i=1,2,\cdots,n)
\een
{\scriptsize 
\ben
\mbox{{\normalsize $\ol{{\bf x}}$}}
&\mbox{{\normalsize $=$}}&
\mbox{{\normalsize {\rm diag}}} 
\left(
-\frac{[\lambda+z_1-z_{n+2}][\lambda+z_{n+1}-z_{n+2}]
[z_1-z_{n+2}]}{[z_1-z_{n+1}] [z_{n+1}-z_{n+2}] [\lambda+z_1-z_{n+2}]},
-\frac{[\lambda+z_2-z_{n+2}][\lambda+z_{n+1}-z_{n+2}][z_2-z_{n+2}]}{[z_2-z_{n+1}][z_{n+1}-z_{n+2}] [\lambda+z_2-z_{n+2}]}, 
\right.  \nn \\ &&  \left. 
\cdots, 
-\frac{[\lambda+z_n-z_{n+2}][\lambda+z_{n+1}-z_{n+2}][z_n-z_{n+2}]}{[z_n-z_{n+1}][z_{n+1}-z_{n+2}] [\lambda+z_n-z_{n+2}]},
\frac{r[\lambda+z_{n+1}-z_{n+2}]}{[z_{n+2}-z_{n+1}][z_{n+1}-z_{n+2}]}
\right)  \nn \\ &&
\left( \ba{ccccc}
1&0&\cdots&0&\disp{-\frac{[z_{n+1}-z_{n+2}][\lambda+z_1-z_{n+2}]}{[\lambda+z_{n+1}-z_{n+2}] [z_1-z_{n+2}]}}\\
0&1&\cdots&0&\disp{-\frac{[z_{n+1}-z_{n+2}] [\lambda+z_2-z_{n+2}]}{[\lambda+z_{n+1}-z_{n+2}][z_2-z_{n+2}]}}\\
\vdots&\vdots&\ddots&\vdots&\vdots \\ 
0&0&\cdots &1&\disp{-\frac{[z_{n+1}-z_{n+2}] [\lambda+z_n-z_{n+2}]}{[\lambda+z_{n+1}-z_{n+2}][z_n-z_{n+2}]}}\\
0&0&\cdots&0&\disp{\frac{[z_{n+1}-z_{n+2}]}{[\lambda+z_{n+1}-z_{n+2}]}}
\ea \right) 
\mbox{{\normalsize ${\bf x}$ \quad $( {\rm if} \quad i=n+1)$}}
\een 
} %scriptsize
\ben
\ol{{\bf x}}={\bf x} && (\mbox{{\rm if}} \quad i=n+2,n+3,\cdots,m-1).
\een 
The birational maps $r_{1,2,\cdots,n+1}$  act on the parameter space as 
\be
r_{1,2,\cdots, n+1}&:&(z_i,\lambda)\mapsto 
(\ol{z_i},-\lambda) \label{olzi}, \quad \mbox{where} \quad
 \ol{z_i} = \left \{ \ba{ll}
z_i+\lambda & (1\leq i \leq n+1) \\ 
z_i  &  (n+2 \leq i \leq m)  
\ea \right. \label{r123} 
\ee
and they act on $\mpp^n$ as
\be
r_{1,\cdots, n+1}&:&(x_0:\cdots:x_n) \mapsto (x_0^{-1}:\cdots:x_n^{-1}).
\ee
Moreover, by $r_{i,i+1}$ and $r_{1,2,\cdots,n+1}$, 
the point $P(z)$ $(z\neq z_i)$ 
is mapped to the point $\ol{P}(z)$:
\be 
\ol{P}(z) &=&
\left(\frac{[\ol{\lambda}+\ol{z}_1-z]}{[\ol{z_1}-z]}:
\cdots:  
\frac{[\ol{\lambda}+\ol{z}_{n+1}-z]}{[\ol{z}_{n+1}-z]} 
\right) , \label{curve3}
\ee
where $(\ol{z}_1,\cdots,\ol{z}_m;\ol{\lambda})$ are given by
$(\ref{r12})$ and $(\ref{r123})$, respectively.
\end{prop}

\bp
Notice that $P_i=(0:\cdots:0:1:0:\cdots:0)$ ($1$ is the $i$-th component).
It is easy to prove for $r_{1,2,\cdots,n+1}$. 
Indeed, by the standard Cremona transformation
with respect to $P_1,\cdots,P_{n+1}$, the point $P(z)$ 
$(z \neq z_i$ $i=1,2,\cdots,n+1)$ 
is mapped to the point
\be 
&&\left(\frac{[z_1-z]}{[\lambda+z_1-z]}:\cdots:  
\frac{[z_{n+1}-z]}{[\lambda+z_{n+1}-z]} \right). 
\label{curve2}
\ee
Suppose (\ref{olzi}), then (\ref{curve2}) coincides with (\ref{curve3}),
while $\ol{P}(\ol{z_i})=P_i(z_i)$ 
is also satisfied for $i=1,2,\cdots,n+1$.
The case of $r_{i,i+1}$ is also shown by the definition of $r_{i,i+1}$ in 
Section~2, normalization of 
$(P_1,\cdots, P_{i+1},P_i,\cdots,P_{n+2})$ to
\ben
&&\left(\ba{c|c} 
& \frac{[\ol{\lambda}+\ol{z}_1-\ol{z}_{n+2}]}{[\ol{z}_1-\ol{z}_{n+2}]} \\
I_{n+1}& \vdots\\
& \frac{[\ol{\lambda}+\ol{z}_{n+1}-\ol{z}_{n+2}]}{[\ol{z}_{n+1}-\ol{z}_{n+2}]}  \ea \right)
\een
and the Riemann relation
\ben [a+b][a-b][c+z][c-z]+ (abc \mbox{ cyclic})&=&0. \een
\ep\\

\begin{remark}
In (\ref{r12}) and (\ref{r123}),  $\lambda\in \mc^{\times}$ 
is an extra-parameter, i.e., $\lambda$ is independent of the parameter
$A$. 
\\
\end{remark}

In order to investigate the preserved quantities for dependent variables,
we present the following conjecture, which has been verified for $n \leq 5$
by M. Eguchi (in a private seminar). 

\begin{conjecture}
Suppose that the blowup points $P_i\in \mpp^n$ lie on an elliptic curve of 
degree $n+1$, then
$\dim|-K_X|=0$ holds if $n$ is even and $\dim|-\frac{1}{2}K_X|=1$ holds 
if $n$ is odd.
\end{conjecture}

This conjecture suggests that our system of odd dimension may reduces to 
$(n-1)$-dimensional system (cf. \cite{takenawa3}).  

\begin{ex}
The translation $t_{\al_0}$ of 
example~\ref{ext} acts on the parameters as
\ben 
z_i &\mapsto& \left\{ \ba{ll}
z_i+9\lambda+4w& (1\leq i \leq n-2)\\
z_i+5\lambda+2w& (i=n-1,n,n+1)\\
z_i+2\lambda+w& (n+2 \leq \leq n+7)\\  
z_i & (n+8 \leq i \leq m)
\ea \right. \\
\lambda &\mapsto& -5\lambda-2w ,
\een
where $w=2z_{n-1}+2z_n+2z_{n+1}-z_{n+2}-\cdots-z_{n+7}$
and acts trivially on any point $P(z)$ on the elliptic curve except on $P_i$ as $z\mapsto z$.

Next, we represent the action of $t_{\al_0}$ on $\mpp^n$ explicitly.
For simplicity, we set $n=4$. One can easily 
generalizes this calculation for $n$-dimensional case.    
In this study,
the actions of $r_{1,2,6,7,8}$ and $r_{1,2,9,10,11}$ are calculated.
Let $i_1,i_2,i_3$ be integers
such that $6\leq i_1<i_2<i_3\leq m$.
By a calculation similar to the one on page 36 of \cite{takenawa3} 
(see also \cite{et}), we have
\ben 
r_{1,2,i_1,i_2,i_3}({\bf x}) 
&=&
\left(\ba{ccccc}
\ol{P}(\ol{z}_1)&\ol{P}(\ol{z}_2)&\ol{P}(\ol{z}_{i_1})&
\ol{P}(\ol{z}_{i_2})&\ol{P}(\ol{z}_{i_3})
\ea \right)
\left(\ba{c}
l_{2,i_1,i_2,i_3}\\
-l_{1,i_1,i_2,i_3}\\
l_{1,2,i_2,i_3}\\
-l_{1,2,i_1,i_3}\\
l_{1,2,i_1,i_2}

\ea \right),
\een
where 
\ben
l_{k_1,k_2,k_3,k_4} &=&
\frac{
\left| \ba{cccc}
P(z)&P(z_{k_1})&\cdots&P(z_{k_4})
\ea \right|
\left| \ba{cccc}
\ol{P}(z)&\ol{P}(\ol{z}_{k_1})&\cdots &\ol{P}(\ol{z}_{k_4})
\ea \right|
}{
\left| \ba{ccccc}
{\bf x}&P(z_{k_1})&\cdots&P(z_{k_4})
\ea \right|},
\een
and $\ol{z}_i$ is given by Prop.\ref{actr}.
\end{ex}

\begin{ex}
The translation $t_{\al_0}$ of the $E_7^{(1)}$ case discussed at the end of 
the 
previous section acts on the parameters as
\ben 
z_i &\mapsto& \left\{ \ba{ll}
z_i+4\lambda+3w& (1\leq i \leq n-3)\\
z_i+3\lambda+2w& (n-2\leq i\leq n+1)\\
z_i+\lambda+w& (n+2 \leq \leq n+5)\\  
z_i & (n+6 \leq i \leq m)
\ea \right. \\
\lambda &\mapsto& -3\lambda-2w ,
\een
where $w=z_{n-2}+z_{n-1}+z_n+z_{n+1}-z_{n+2}-\cdots-z_{n+5}$
and acts trivially on any point $P(z)$ on the elliptic curve except on $P_i$.
\end{ex}

\medskip

\section{Conclusion}

In this paper, we have proposed $n$-dimensional dynamical systems 
associated with
translations of affine Weyl groups, which are included in a Weyl group of
indefinite type. In order to examine their integrability, we
computed their algebraic entropy and investigated the existence of preserved
divisor classes.
We also studied the case where the points of blowing-up 
are on some elliptic curve. Following is a comparison of our systems with
the elliptic discrete Painlev\'e equation.

\noi $\cdot$\
The elliptic discrete Painlev\'e is an isomorphism of a family of rational
surfaces. Its degree grows in the quadratic order. The evolution of
the parameters can be written by elliptic functions.

\noi $\cdot$\
Our system in general case is a pseudo-isomorphism of a family of rational
varieties. Although its degree grows in the quadratic order, the evolution of 
the parameter may not be able to be solved.

\noi $\cdot$\
Our system in the elliptic case is a pseudo-isomorphism of a family of rational
varieties. Its degree grows in the quadratic order, and the evolution of 
the parameter can be written by elliptic functions. Moreover, we have 
conjectured that the odd-dimensional system reduces to 
$(n-1)$-dimensional one. 

\noi $\cdot$\
For further comparison, we give some properties of dynamical systems
associated with a general element of infinite order of the Weyl group
$W(n,m)$. Such system is a pseudo-isomorphism of a family of rational
varieties. Its degree grows exponentially, i.e. the algebraic entropy is 
positive. The evolution of the parameter may not be able to be solved.\\

{\noindent{\it Acknowledgment.}}
The author would like to thank M.Eguchi, T. Masuda and M. Noumi 
for their suggestion
on the case where the points of the parameters lie on an elliptic curve
and the members of the Okamoto-Sakai seminar 
at the Univ. of Tokyo and the Kyusyu integrable system seminar
for discussions. The author also thanks the anonymous referees 
for their useful comments and suggestions.
Moreover, the assistance from the Sumitomo foundation and the
Japan Society for the Promotion of Science are appreciated.


\begin{thebibliography}{99}

\bibitem{bk}
E. Bedford and K. Kim, {On the degree growth of birational mappings
in higher dimension}, 
{\it J. Geom. Anal.} \textbf{14} (2004), 567--596.


\bibitem{bellon}
M. P. Bellon,
{Algebraic entropy of birational maps with invariant curves}, 
{\it Lett. Math. Phys.} \textbf{50} (1999), 79--90.

\bibitem{bhm}
S. Boukraa, S. Hassani and J. M. Maillard, 
{Noetherian mappings}
{\it Phys. D} \textbf{185} (2003), 3--44


\bibitem{cantat}
S. Cantat,
{Dynamique des automorphismes des surfaces projectives complexes}, 
{\it C. R. Acad. Sci. Paris Ser. I Math.} \textbf{328} (1999), 901--906.


\bibitem{coble}
A. B. Coble, {\it Algebraic geometry and theta functions. 
American Mathematical Society Colloquium Publications, vol. X}, 
American Mathematical Society, Providence, 1929


\bibitem{do}
I. Dolgachev and D. Ortland, {\it Point sets in projective
spaces and theta functions},
Ast\'{e}risque Soc.Math.de France \textbf{165}, 1988

\bibitem{df}
J. Diller and C. Favre, 
{Dynamics of bimeromorphic maps of surfaces},
{\it Amer. J. Math.} \textbf{123} (2001), 1135--1169  

\bibitem{et}
M. Eguchi and T. Takenawa, 
{Elliptic curves and birational representation of Weyl groups},
{\it Preprint} arXiv:math/0511726   


\bibitem{grp}
B. Grammaticos, A. Ramani and V. Papageorgiou, {Do integrable mappings
  have the Painlev\'{e} property?},
{\it Phys. Rev. Lett.} \textbf{67} (1991), 1825--1827 

\bibitem{gromov}
M. Gromov, 
{On the entropy of holomorphic maps}, {\it Enseign. Math.} \textbf{49} 
(2003), 217--235


\bibitem{hv} J. Hietarinta and C. M. Viallet, {Singularity confinement
and chaos in discrete systems}, {\it Phys. Rev. Lett.} \textbf{81} (1997),
325--328

\bibitem{kac}
V. Kac, {\it Infinite dimensional lie algebras, 3rd ed.},
Cambridge University Press, Cambridge, 1990

\bibitem{kmnoy}
K. Kajiwara, T. Masuda, M. Noumi, Y. Ohta and Y. Yamada,
{${}\sb {10}E\sb 9$ solution to the elliptic Painleve equation},
 {\it J. Phys. A} \textbf{36} (2003),  L263--L272\\
{Point configurations, Cremona transformations and the elliptic 
difference Painleve equation} {\it Preprint} nlin.SI/0411003

\bibitem{looijenga}
E. Looijenga, {Rational surfaces with an anti-canonical cycle},
{\it Annals of Math.} \textbf{114} (1981), 267--322

\bibitem{rgh}
A. Ramani, B. Grammaticos and J. Hietarinta, {Discrete versions of the
  Painlev\'{e} equations},
{\it Phys. Rev. Lett.} \textbf{67} (1991), 1829--1832

\bibitem{su}
M. H. Saito and H. Umemura, 
{Painlev\'e equations and deformations of rational surfaces 
with rational double points},
{\it Physics and combinatorics 1999 (Nagoya)}, 320--365, 
World Sci. Publishing, River Edge, 2001


\bibitem{sakai} H. Sakai,
{Rational surfaces associated with affine root systems and geometry of
the Painlev\'{e} equations}, {\it Commun. Math. Phys.} {\bf 220} (2001),
165--229

\bibitem{sibony} N. Sibony, 
{\it Dynamique des applications rationnelles de ${\mathbb P}^k$}
in {\it Panor. Synths\`es, 8}, 97--185, Soc. Math. France, Paris, 1999


\bibitem{takenawa} T. Takenawa, {Discrete dynamical systems associated
with root systems of indefinite type}, {\it Commun. Math. Phys.} 
\textbf{224} (2001), 657--681

\bibitem{takenawa2}
T. Takenawa,
{Algebraic entropy and the space of initial values for discrete
dynamical systems}, {\it J. Phys. A: Math. Gen.} 
\textbf{34} (2001), 10533--10545

\bibitem{takenawa3} T. Takenawa, {Discrete dynamical systems associated 
with the configuration space of 8 points in $\mpp^3(\mc)$}, 
{\it Commun. Math. Phys.} \textbf{246} (2004), 19--42

\bibitem{takenawaet}
T. Takenawa, M. Eguchi, B. Grammaticos, Y. Ohta, A. Ramani and 
J. Satsuma,
{Space of initial conditions for linearizable mappings}, {\it Nonlinearity}
\textbf{16} (2003), 457--477

\bibitem{yomdin}
Y. Yomdin, {Volume growth and entropy}, {\it Israel J. Math}
\textbf{57} (1987), 285--300

\end{thebibliography}
\end{document}